\documentclass[journal]{IEEEtran} 
\usepackage{cite}
\usepackage{graphicx,color}
\usepackage{amsfonts, amssymb, amsmath, mathrsfs}
\usepackage{enumerate}
\usepackage{changepage}
\usepackage{subfig}

\definecolor{dred}{rgb}{0.8,0,0} 
\definecolor{dgreen}{rgb}{0,0.8,0} 
\definecolor{dblue}{rgb}{0,0,0.8} 
\definecolor{dyellow}{rgb}{0.4,0.4,0} 
\definecolor{dcyan}{rgb}{0,0.4,0.4} 
\definecolor{dpurple}{rgb}{0.4,0,0.4} 

\begin{document}

\title{Impact of Rocks and Minerals on Underground Magneto-Inductive Communication and Localization}

\author{Traian~E.~Abrudan,~\IEEEmembership{Member,~IEEE,}
        Orfeas~Kypris,~\IEEEmembership{Member,~IEEE,}
        Niki~Trigoni,
        and~Andrew~Markham
\thanks{The authors would like to thank EPSRC for funding this research 
(Grant ref. EP/L00416X/1 Digital Personhood: Being There: Humans and Robots in Public Spaces (HARPS), and
Grant ref. EP/M017583/1 Magneto-Inductive Six Degree of Freedom Smart Sensors (MiSixthSense) for Structural and Ground Health Monitoring),}
\thanks{Manuscript received Xxx xx, 2016; revised  Xxx xx, 2016.}}

\markboth{~}%
{Abrudan \MakeLowercase{\textit{et al.}}: }

\maketitle

\begin{abstract}
In this paper, we analyze the effect of different underground materials on very-low and low frequency magnetic fields used in the contexts of magneto-inductive localization and communication applications, respectively.
We calculate the attenuation that these magnetic fields are subject to while passing through most common rocks and minerals. 
Knowing the attenuation properties is crucial in the design of underground magneto-inductive communication systems.
In addition, we provide means to predict the distortions in the magnetic field that impair localization systems. 
The proposed work offers basic design guidelines for communication and localization systems in terms of channel path-loss, operation frequencies and bandwidth. 
For the sake of the reproducibility of the results, we provide the raw data and processing source code to be used by the two research communities.
\end{abstract}


\IEEEpeerreviewmaketitle

\section{Introduction}

\IEEEPARstart{T}{he} motivation of this paper stems from two different underground applications of very low frequency magnetic fields:
Magneto-Inductive (MI) {\em communications} \cite{GabDegWai71,SunAky10,SojWraDin01,MarTri12,KisAkyGer14,AkySunVur09,TanSunAky15} and
MI {\em localization} \cite{AbrXiaMarTri16TGRS,AbrXiaMarTri15JSAC,MarTriEllMac12,MarTriEllMac10,DavCheTucAtk08,SogVicVan_etal04}.
Both research communities may benefit from the results provided here for system design, without requiring the tedious work of searching for values of electrical constants for different underground materials. 
We evaluate the attenuation that magnetic fields experience at three very different operating frequencies: 1~kHz, 100~kHz and 10~MHz, respectively.
The kHz range is typically used for underground localization, whereas for communication, a higher carrier frequency is used. 
On one hand, lower frequencies (few kHz) penetrate deeply into most natural underground materials, and suffer from low environmental distortions. Therefore, they are useful for through-the-earth magneto-inductive localization. 
On the other hand, lower frequencies do not allow for a wide signal bandwidth, and therefore, they are unable to carry much information. 
For this reason, wireless underground sensor networks~\cite{AkyStu06} use frequencies a few order of magnitudes higher (typically in the MHz range).
However, higher frequencies experience much higher attenuation in conductive materials due to the skin effect. 
Therefore, being able to choose an appropriate operating frequency for the application at hand  is crucial, and requires understanding the nature of the underground medium. 

In {\em MI communications}, the most important metric is the channel capacity, which depends on bandwidth and signal-to-noise ratio (SNR) via the well-known Shannon capacity expression. 
Channel bandwidth is determined by the frequency response of transmission medium, as well as the quality factors of the resonant coil antennas, whereas SNR is dictated by the field attenuation through the medium, ambient noise, and receiver sensitivity. 
Operation frequency should be chosen according to the nature of the transmission medium in order to achieve the highest SNR, and obtain sufficient bandwidth. Higher carrier frequencies allow for larger bandwidth, and therefore, higher data rates.
Overall, the capacity of a MI communication system is optimized both in terms of hardware design, and network architecture.
The hardware optimization mainly includes the design of coil antennas or waveguides~\cite{Gib10,SunAky10,TanSunAky15}, whereas the architecture addresses
the underlying signal processing techniques \cite{AkyStu06,LinAkyWanSun15,PriHow04,AkySunVur09,AbrXiaMarTri16TGRS,AbrXiaMarTri15JSAC}, medium multiple access to minimize interference \cite{KisAkyGer14,SojWraDin01,LiVurAky07,MarTri12}, and topology, etc. \cite{LinAkyWanSun15}. 

In {\em MI localization}, the requirements are even stricter than in MI communications. 
Not only is attenuation important, but also preserving the shape of the generated magnetic field.
Often, MI localization relies on the entire vector field (not just its magnitude), either by exploiting its dipole shape \cite{AbrXiaMarTri15JSAC,AbrXiaMarTri16TGRS}, or other geometrical properties such as null-field \cite{DavCheTucAtk08,SogVicVan_etal04,AyuCucLerVil10}.
Therefore, in order to achieve good positioning accuracy, one must either operate in the quasi-static region, whose limit is dictated by the medium characteristics, or ensure that the distortions do not affect the desired geometrical properties of the field beyond the quasi-static limit \cite{DavCheTucAtk08,SogVicVan_etal04}. 

In this paper, we provide attenuation figures for most common underground materials at the three different frequencies typically used in communications and localization, respectively. 
We attempt to help researchers working in these topics to find answers to fundamental questions such as: 
Given a set of transmitter/receiver parameters, can we communicate through a thick layer of granite or limestone? 
If so, how far can we communicate? 
What would be a good operation frequency for an underwater wireless sensor network? 
Is a magneto-inductive localization system that relies on the dipole equations capable to achieve reasonable accuracy in a calcite mine, or in a salt mine? 
What type of minerals are detrimental to my system? 
We make the following contributions:
\begin{enumerate}
\item We provide an extensive survey of electromagnetic properties of most common underground materials from multiple sources:
tabulated numerical values and ranges of electrical resistivity, electrical permittivity and magnetic permeability~\cite{Mathworks_underground};
\item We give a comprehensive classification of rocks and minerals in terms of attenuation experienced by magnetic fields at different frequencies;
\item We devise basic design parameter values such as path-loss, optimal operation frequencies and bandwidth, that help researchers working on MI communications and MI localization to predict the operation of their system in certain underground environments.
\end{enumerate}

\section{Underground Transmission Medium}
The underground magnetic transmission medium consists mostly of inorganic materials (rock, soil, minerals, water and gases, etc.), and rarely, organic materials. 
The vast majority of natural underground materials have relative magnetic permeability close to the free-space value, and therefore permeability does not play a crucial role in the field characterization. 
However, electrical conductivity and permittivity of underground materials depend heavily on water content, chemical composition and constitution, as well as environmental factors such as temperature, pressure, etc.
Typically, the sub-surface structure is stratified, each layer having different thickness and electromagnetic characteristics, 
and therefore, it is non-trivial to characterize, although stratified earth models exist \cite{Wai71}.
For the layered structure, the concept of ``effective parameters'' has been introduced \cite{ITU-R_R527}, which enable the use of homogeneous ground models.
There are also various scenarios, such as mines, where there is a dominant layer of material whose properties dictate the values of the effective parameters.

Due to its quasi-static nature, the very low frequency magnetic field can be modeled in free-space using the magnetic dipole equations \cite{Bla96}. 
The magnitude and phase of magnetic fields is affected by two key material properties: {\em magnetic permeability}, $\mu$ and {\em electrical conductivity}, $\sigma$. These may undermine the validity of the dipole model underground.
Most rocks are non-magnetic, and magnetic mineral are surprisingly few \cite{TelGelShe90}.
Conductivity of the materials gives rise to eddy currents that produce an out-of-phase secondary field \cite{SogVicVan_etal04}. 
This secondary field superimposes with the primary field, thus distorting the dipole field shape. 
High conductivity also leads to higher attenuation of the field magnitude that passes through. 
There is a also frequency dependence of the electromagnetic constants, but it only becomes relevant at frequencies higher than the ones considered in this paper (see Fig 1 in~\cite{ITU-R_R527}). In the next section, we address material electromagnetic properties in more detail.

\subsection{Magnetic Permeability}
In this section, we address the material electromagnetic properties that are relevant in the contexts of underground positioning and underground communication using very low frequency magnetic fields.
Magnetic permeability $\mu$ 
quantifies the extent of magnetization that a material obtains in the presence of an external magnetic field. 
It is denoted by $\mu = \mu_0 \mu_{\text{r}}$, where $\mu_0$ is the permeability of free space, and $\mu_{\text{r}}$ is the relative permeability, which varies depending on the type of material. 
Magnetic materials typically possess permeability values that differ from free-space, thus changing the direction of the incident vector field at the interface between non-magnetic and magnetic media,
and invalidating the dipole equations. However, Telford {\em et al.}~\cite{TelGelShe90} address in detail the magnetism of rocks and minerals; 
they show that the vast majority of rocks are non-magnetic, and that magnetically important minerals are surprisingly few in number~\cite[Sec. {3.3.5}]{TelGelShe90}.
They also point out that ferromagnetic materials do not exist in nature, and that, practically, all minerals are ferrimagnetic. 
A table with magnetic permeability for some common minerals is provided in~\cite[p. 291]{TelGelShe90}, and except for magnetite (an iron oxide mineral), pyrrhotite (an iron sulfide mineral) and titanomagnetite, (a mineral of the complex oxide class), all the other minerals have $\mu_{\text{r}} \approx 1$.
Magnetic susceptibility $\chi=\mu_{\text{r}}-1$, is another measure which quantifies the extent to which a material can be magnetized by an external field. 
Table 3.1 (page 74) in \cite{TelGelShe90} provides a very detailed description of magnetic susceptibility for different rocks and minerals (ranges, and few average values).
Most common minerals such as coal, rock salt, graphite, quartz, gypsum, calcite, clay, etc. have magnetic susceptibility close to zero (in general, below $10^{-3}$), 
hence, the magnetic susceptibility is too small to change the relative permeability appreciably from unity~\cite[Sec. {3.4.3}]{TelGelShe90}.
The same can be said about the most common types of sedimentary rocks such as dolomite, limestone, sandstone, etc., metamorphic rocks such as amphibolite, schist, phyllite, quartzite, etc.,
and igneous rocks such as granite, rhiolyte, dolorite, basalts, andesite, etc.
In conclusion, since most soils in nature do not contain massive amounts of magnetic minerals in high concentrations, 
we can safely assume that the relative magnetic permeability 
of most underground environments is close to one, 
as also pointed out in~\cite{ITU-R_R527,Gib10}. 
The same assumption was used in \cite{SunAky10} for the magneto-inductive communications. 

\subsection{Electric Permittivity and Electrical Conductivity}

Electric permittivity is a complex quantity and defined as $\varepsilon = \varepsilon' - \jmath \varepsilon''$, where $\varepsilon'$ and $\varepsilon''$ are the real and imaginary parts respectively. The real part quantifies the polarizability of a dielectric medium subjected to an external electric field, and at frequencies higher than the ones addressed in the present paper gives rise to a \emph{displacement} current. The imaginary part quantifies the dissipation of energy into heat and is closely related to the electrical conductivity $\sigma$, since it gives rise to the \emph{conduction} current. The ratio between the two parts corresponds to the phase lag between electric and magnetic fields. A commonly used expression for the relative electric permittivity is:
\begin{equation}
\varepsilon_{\text{r}} = \varepsilon_{\text{r}}' - \frac{\jmath\sigma}{\omega \varepsilon_0}
\label{eqn:permittivity}
\end{equation}
Strictly speaking, $\varepsilon_{\text{r}}'$ and $\sigma$ are both also functions of frequency; for the frequency range we are considering in the present paper, however, it is safe to use the DC values~\cite{ITU-R_R527}. As seen in \eqref{eqn:permittivity}, the imaginary part of the permittivity arises due to the conductivity.
Like the electrical conductivity, the real part of the relative permittivity $\varepsilon_{\text{r}}$ varies with the water content (since water has $\varepsilon_{\text{r}} \approx 80$).
Unlike relative permitivity that usually exhibits values that ranges from few units to few tens, the conductivity values
may easily differ by many orders of magnitude depending on the water content. 
For example, dry to moist clay has a typical range of $\varepsilon_{\text{r}}=7,\ldots,43$~\cite[Sec. {5.4.2}]{TelGelShe90}. 
Therefore, the influence of electric permittivity on skin depth remains very low compared to the impact of conductivity. 

Electrical conductivity $\sigma$ gives rise to energy dissipation in the material due to eddy currents that produce an out-of-phase secondary field \cite{SogVicVan_etal04}. 
This field adds to the primary field, thus distorting its shape. 
As a result, the field magnitude decays fast through the material. The decay is exponential and is associated with the skin effect~\cite{DavCheTucAtk08}. 
Electrical conduction through the ground takes three different forms: electronic (ohmic), electrolytic (ionic), and dielectric (due to polarization). 
Dry rocks exhibit very low conductivity, but porous rocks can absorb large quantities of mineralized water, thus increasing conductivity up to 80 times~\cite{Gib10}.
Igneous rock tend to have the lowest conductivity, whereas sedimentary rocks have the highest \cite{TelGelShe90}, 
but conductivity varies with the age of the rock, location and local conditions~\cite{Gib10}. 
However, the conductivity of most underground materials is sufficiently low, such that eddy currents can be ignored at very low frequencies~\cite{SogVicVan_etal04,DavCheTucAtk08}. 
By contrast, high frequency radio waves experience extreme attenuation through the conductive ground, as well as distortion due to reflections.

\section{Impact of Electromagnetic Properties on MI Communication and Localization}

\subsection{Attenuation}

In an infinite medium, the magnetic flux density $B$ at distance $r$ is related to the flux density $B_0$ at the origin by the expression
\begin{equation}
B(r) = B_0 e^{-\jmath k r}
\end{equation}
where $r$ is distance, and 
\begin{equation}
k = k' - \jmath k''= \omega \sqrt{\mu \varepsilon},
\label{eqn:wavenumber}
\end{equation}
is the complex wavenumber. The real part, $k'$, is inversely proportional to wavelength $\lambda$ such that $\lambda = 2\pi/k'$, and is responsible for the oscillatory behaviour of the field. This is also commonly referred to as $\beta$, termed the \emph{propagation constant}.
The imaginary part, $k''$, is responsible for attenuation over distance. This is also commonly referred to as $\alpha$, termed the \emph{attenuation constant}. Its reciprocal (i.e. $1/\alpha$) is referred to as the \emph{skin depth} $\delta$, which is the distance $r$ at which $B(r)/B_0 = 1/e$ (approx. 8.7~dB.). A general expression for the skin depth, obtained by substituting \eqref{eqn:permittivity} into \eqref{eqn:wavenumber} and then separating real and imaginary parts, is the following~\cite{JorBal68},
\begin{equation}
\delta = \left[\omega \sqrt{\frac{\mu \varepsilon'}{2} \left(  \sqrt{1 + \Big[\frac{\sigma}{\omega \varepsilon'}\Big]^2} -1  \right)}~\right]^{-1}
\label{eqn::delta}
\end{equation}
which is also what we used in our calculations. When electrical conductivity is the dominating property such that $\varepsilon'\omega/\sigma \ll 1$ and the considered materials are ``good'' conductors, then $k'=k''$ and the skin depth reduces to the well-known expression $\delta=\sqrt{2/\sigma \omega \mu}$. 

The attenuation of 8.7~dB per skin depth corresponds to plane wave. 
For an induction loop antenna, the overall path loss consists of three terms~\cite{Gib10} that depend on the distance as follows:
i) the inverse cube attenuation term (i.e., 60~dB/decade); ii) the exponential attenuation term corresponding to skin effect; iii) a skin depth dependent extra-term that {\em reduces} the effect of ii).
Therefore, the skin depth distance is a lower bound on field penetration distance, and it is safe to rely on it. 

In the frequency domain, the underground transmission medium exhibits a band-pass behavior~\cite{Gib10}. 
At the lower edge of the band, the Faraday's law dominates (higher current is induced at higher frequencies), whereas at the upper edge of the band, the skin effect comes into play (attenuation increases with frequency). 
Therefore, there is a critical frequency where attenuation is minimized and that depends on the material properties. 
In Section~\ref{sec:design_guidelines}, we provide attenuation figures for the most common rocks and minerals, as well as the optimal operation frequency for magneto-inductive communications, which corresponds to minimum attenuation.

\subsection{Near-Field and Quasi-Static Boundaries} 
There are various definitions of the near field region in the literature. In \cite{Agb11}, more than ten different definitions for the free-space near/far-field boundary are summarized, 
depending on wavelength, antenna aperture, etc. Therefore, such a boundary is rather a matter of convention, depending on antenna type (electrical antenna of induction loop), operation frequency, 
and properties of the transmission medium (the wavelength might not be the same as in free-space). 
Near or induction field corresponds to the region where there is no significant radiation due to the fact that electric and magnetic fields are in quadrature, 
whereas far or radiation field corresponds to the region where the electric and magnetic field are in phase, and therefore, convey energy~\cite{Gib10}.
For electromagnetically short antennas (shorter than half of the wavelength of emitted radiation), definitions for the near-field, transition zone, and the far-field are listed in Table \ref{table_definitions}.

\begin{table}[]
\centering
\begin{tabular}{l|lll}
\hline
Properties & Near field       & Transition zone           & Far Field         \\ \hline
wavelength & $ r \ll \lambda$ & $\lambda < r,< 2 \lambda$ & $ r \gg 2\lambda$ \\
wavenumber & $|kr| \ll 2\pi$  & $2\pi < |kr|  < 4\pi$     & $ |kr| \gg 4\pi$ \\
skin depth & $r\ll 2\pi\delta$  & $2\pi\delta< r <4\pi\delta$     & $r\gg 2\pi\delta$ \\
\hline
\end{tabular}
\caption{Equivalent definitions of near-field, transition zone, and far-field using three common measures (wavelength $\lambda$, wavenumber $|k|$, and skin depth $\delta$) for electromagnetically short antennas.}
\label{table_definitions}
\end{table}

Next, we provide skin depth values for the most common rocks and mineral, and this can be directly use to determine the quasi-static region.


\section{Design Guidelines for MI Localization and Communication} \label{sec:design_guidelines}

\subsection{Skin Depth Values in Rocks and Minerals} \label{sec:skin_depth}

In order to quantify the achievable MI communication range, as well as the feasible operation distance for magneto-inductive localization (the quasi-static region),
we provide a comprehensive evaluation of skin depth at three different frequencies. 
The following assumptions were made: 
1) We used the electromagnetic constants (electric conductivity, dielectric constant, and magnetic permeability) provided in~\cite{TelGelShe90, 
Par67,
Cla66,
BirSchSpi50,
NRC_etal_27,
RosSmi36,
Shu12,
Pat15,
Spe05,
Ves91,
KniNur97,
AhmZih90,
Gai14,
GuiPla_etal10,
Ito15,
LowMar70,
KatRab61,
Mad04,
OsoMgb14,
Mar_12,
Lin16,
CliCon16,
Vega16} for the most common underground materials,
and for usual underground temperature ranges and different water content (dry to moist); 
For the sake of the reproducibility of the results, the raw values of all electrical constants used in this paper along with the corresponding source codes for data processing are available at~\cite{Mathworks_underground}.
2) For conductivity, we used minimum, maximum values (and average value, if available);
3) For permittivity we only used minimum value (and maximum value, if available);
4) Relative magnetic permeabilities were replaced by a typical value;
5) We assumed that the electromagnetic constants do not vary significantly with frequency at low frequencies, as demonstrated in Fig 1 in~\cite{ITU-R_R527} for different soils, mineralized water and ice;

\begin{figure*}[!h]
  \begin{center}
    \includegraphics[clip, width=\textwidth]{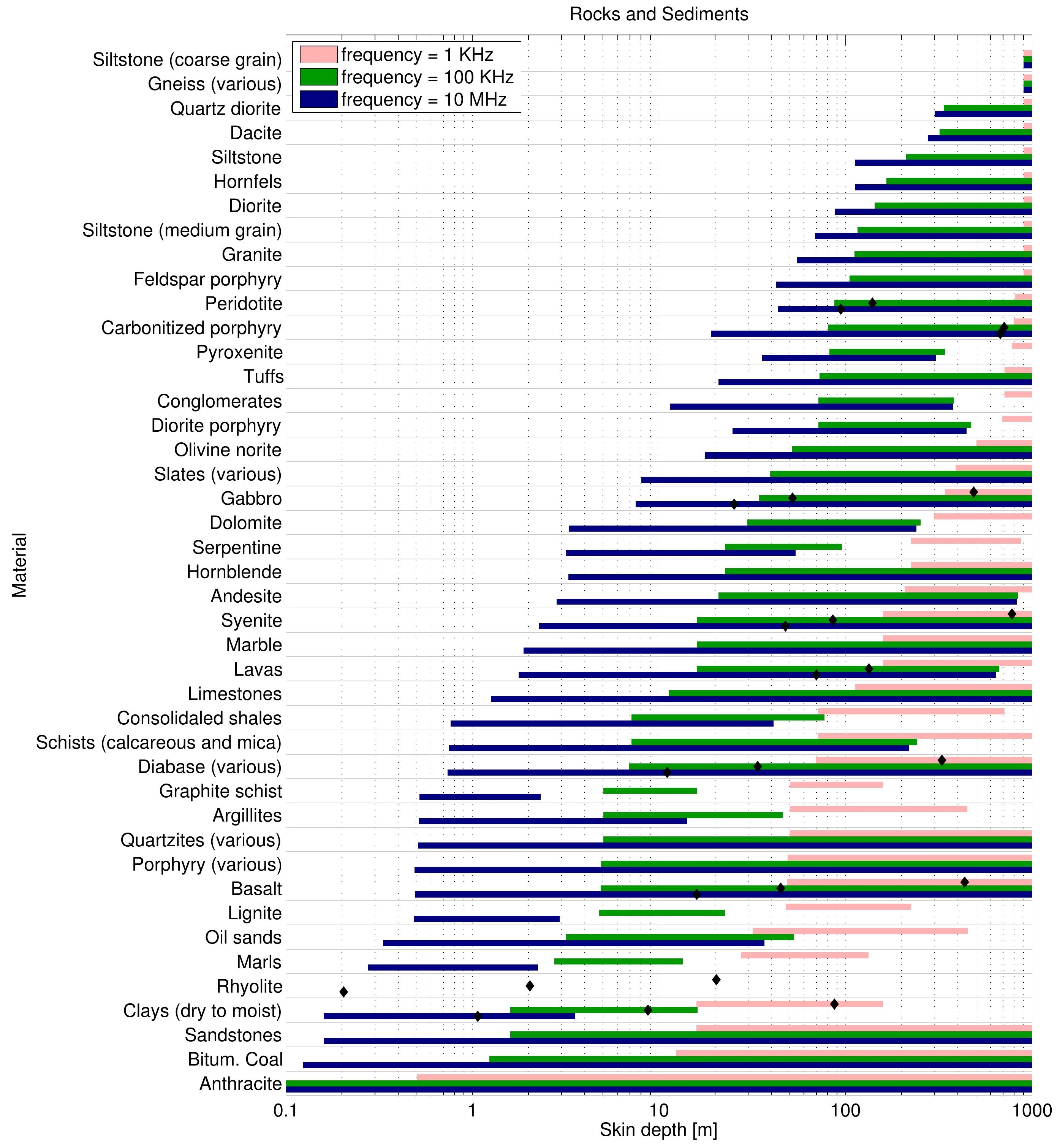}
    \caption{Bar plot of skin depth for different rocks and sediments at three different frequencies: 1~kHz, 100~kHz and 10~MHz (sorted in descending order, by minimum value). The minimum skin depth value corresponding to the 2 top materials exceeds 1~km.
     The average skin depth value (where available) is indicated by the black diamond. \label{fig_skin_rocks}}
    \end{center}
\end{figure*}

\begin{figure*}[!h]
  \begin{center}
    \includegraphics[clip, width=\textwidth]{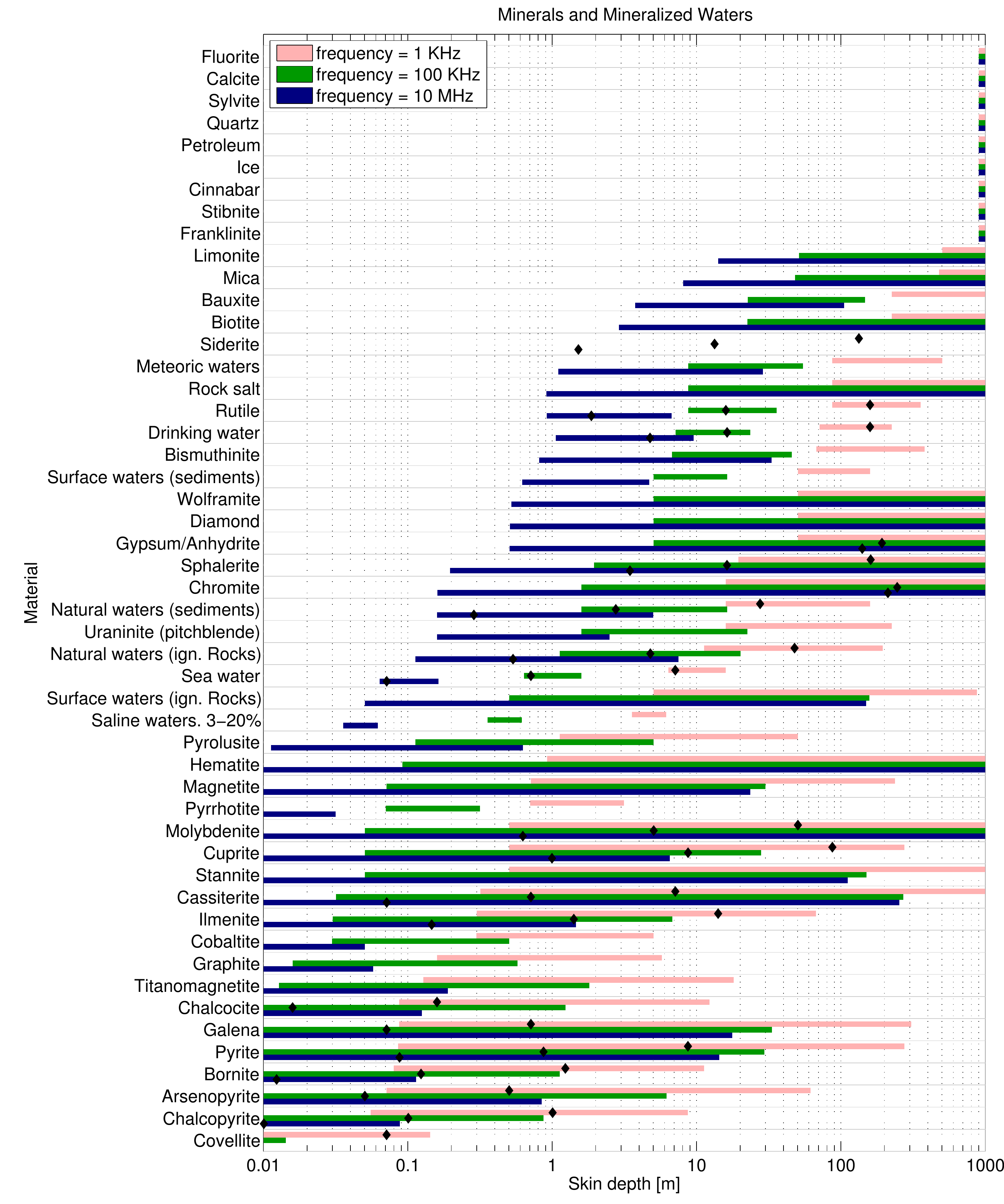}
    \caption{Bar plot of skin depth for different minerals and mineralized waters at three different frequencies: 1~kHz, 100~kHz and 10~MHz (sorted in descending order, by minimum value). The minimum skin depth value corresponding to the 9 top materials exceeds 1~km.
     The average skin depth value (where available) is indicated by the black diamond. \label{fig_skin_minerals}}
    \end{center}
\end{figure*}

In Tables \ref{table_skin_rocks} and \ref{table_skin_minerals}, we provide values of the skin depth for most common underground materials. 
For an easier visualization, we provide bar plots of the skin depth for different rocks and sediments in Fig.~\ref{fig_skin_rocks}, and for different minerals and ores in Fig.~\ref{fig_skin_minerals}, 
at frequencies of 1~kHz, 100~kHz, and 10~MHz (sorted in descending order, by minimum value).
We may notice in Fig~\ref{fig_skin_rocks} and Fig.~\ref{fig_skin_minerals} that the 1~kHz magnetic field penetrates deeply into most rocks, sediments and water, except for some highly conductive coals, and heavily mineralized waters (electrolytes).
The same is valid for most common minerals and ores, except for a few highly conductive sulfides of metals and graphite, as also shown in Table \ref{table_skin_minerals}. 
By contrast, the penetration capabilities of 10~MHz magnetic field are substantially diminished.

\begin{table*}[!h]
{\small
\begin{center}
\begin{tabular}{|p{6cm}||c|c|c|c|}
\hline 
    {\bf Rocks and sediments} & \multicolumn{3}{|c|}{\bf Skin depth [m]} & {\bf References} \\
    \cline{2-4}
    ~ & {\bf (f = 1~kHz)} & {\bf (f = 100~kHz)} & {\bf (f = 10~MHz)} & ~ \\
\hline\hline
     Siltstone (coarse grain)   &   $\geq$1000   &   $\geq$1000   &   $\geq$1000   &     \cite{TelGelShe90,Ves91}   \\
    Gneiss (various)   &   $\geq$1000   &   $\geq$1000   &   $\geq$1000   &     \cite{TelGelShe90,BirSchSpi50}   \\
    Quartz diorite   &   $\geq$1000   &   $\geq$336.03   &   $\geq$300.32   &     \cite{TelGelShe90}   \\
    Dacite   &   $\geq$1000   &   $\geq$319.15   &   $\geq$276.88   &     \cite{TelGelShe90}   \\
    Siltstone   &   $\geq$1000   &   $\geq$211.81   &   $\geq$112.67   &     \cite{TelGelShe90,Par67}   \\
    Hornfels   &   $\geq$1000   &   $\geq$165.95   &   $\geq$112.38   &     \cite{TelGelShe90,BirSchSpi50}   \\
    Diorite   &   $\geq$1000   &   $\geq$143.53   &   $\geq$87.41   &     \cite{TelGelShe90}   \\
    Siltstone (medium grain)   &   $\geq$1000   &   $\geq$116.13   &   $\geq$68.5   &     \cite{TelGelShe90,Ves91}   \\
    Granite   &   $\geq$1000   &   $\geq$111.84   &   $\geq$55.01   &     \cite{TelGelShe90,Par67}   \\
    Feldspar porphyry   &   $\geq$1000   &   $\geq$105.23   &   $\geq$42.54   &     \cite{TelGelShe90}   \\
    Peridotite   &   $\geq$813.47   &   $\geq$87.32   &   $\geq$43.58   &     \cite{TelGelShe90}   \\
    Carbonitized porphyry   &   $\geq$795.89   &   $\geq$80.69   &   $\geq$19.06   &     \cite{TelGelShe90}   \\
    Pyroxenite   &   $\geq$780.11   &   82.17--340.45   &   35.8--304.58   &     \cite{TelGelShe90}   \\
    Tuffs   &   $\geq$711.91   &   $\geq$72.7   &   $\geq$20.84   &     \cite{TelGelShe90,Cla66}   \\
    Conglomerates   &   $\geq$711.8   &   71.57--381.23   &   11.5--375.39   &     \cite{TelGelShe90}   \\
    Diorite porphyry   &   $\geq$693.96   &   71.61--470.71   &   24.78--445.95   &     \cite{TelGelShe90,Par67,BirSchSpi50}   \\
    Olivine norite   &   $\geq$503.44   &   $\geq$51.88   &   $\geq$17.59   &     \cite{TelGelShe90,Par67}   \\
    Slates (various)   &   $\geq$389.89   &   $\geq$39.38   &   $\geq$8.03   &     \cite{TelGelShe90,Mar_12,NRC_etal_27}   \\
    Gabbro   &   $\geq$340.63   &   $\geq$34.46   &   $\geq$7.49   &     \cite{TelGelShe90}   \\
    Dolomite   &   $\geq$297.75   &   29.8--253.23   &   3.28--240.25   &     \cite{TelGelShe90,Par67}   \\
    Serpentine   &   225.09--872.56   &   22.59--95.88   &   3.16--53.98   &     \cite{TelGelShe90}   \\
    Hornblende   &   $\geq$225.07   &   $\geq$22.6   &   $\geq$3.27   &     \cite{TelGelShe90}   \\
    Andesite   &   $\geq$207.52   &   20.82--838.65   &   2.83--827.57   &     \cite{TelGelShe90,Par67}   \\
    Syenite   &   $\geq$159.16   &   $\geq$15.98   &   $\geq$2.28   &     \cite{TelGelShe90}   \\
    Marble   &   $\geq$159.16   &   $\geq$15.94   &   $\geq$1.88   &     \cite{TelGelShe90,Par67}   \\
    Lavas   &   $\geq$159.16   &   15.93--666.88   &   1.77--639.27   &     \cite{TelGelShe90,BirSchSpi50}   \\
    Limestones   &   $\geq$112.54   &   $\geq$11.27   &   $\geq$1.25   &     \cite{TelGelShe90,Par67,DavCheTucAtk08}   \\
    Consolidaled shales   &   71.18--712.36   &   7.12--77.34   &   0.76--41.14   &     \cite{TelGelShe90,KniNur97}   \\
    Schists (calcareous and mica)   &   $\geq$71.18   &   7.12--242.53   &   0.75--218.89   &     \cite{TelGelShe90,Par67,BirSchSpi50}   \\
    Diabase (various)   &   $\geq$69.3   &   $\geq$6.93   &   $\geq$0.73   &     \cite{TelGelShe90}   \\
    Graphite schist   &   50.33--159.16   &   5.03--15.98   &   0.52--2.33   &     \cite{TelGelShe90,CliCon16}   \\
    Argillites   &   50.33--450.27   &   5.03--46.13   &   0.51--14.16   &     \cite{TelGelShe90,Ves91}   \\
    Quartzites (various)   &   $\geq$50.33   &   $\geq$5.03   &   $\geq$0.51   &     \cite{TelGelShe90,Par67,BirSchSpi50}   \\
    Porphyry (various)   &   $\geq$48.88   &   $\geq$4.89   &   $\geq$0.49   &     \cite{TelGelShe90}   \\
    Basalt   &   $\geq$48.66   &   $\geq$4.87   &   $\geq$0.49   &     \cite{TelGelShe90,AhmZih90,Pat15}   \\
    Lignite   &   47.75--225.09   &   4.78--22.57   &   0.48--2.94   &     \cite{TelGelShe90,Spe05}   \\
    Oil sands   &   31.83--450.91   &   3.18--53.04   &   0.33--36.78   &     \cite{TelGelShe90,Gai14}   \\
    Marls   &   27.57--133.17   &   2.76--13.4   &   0.28--2.25   &     \cite{TelGelShe90,Par67}   \\
    Rhyolite   &   20.32--20.32   &   2.03--2.03   &   0.2--0.2   &     \cite{TelGelShe90}   \\
    Clays (dry to moist)   &   15.92--159.17   &   1.59--16.11   &   0.16--3.55   &     \cite{TelGelShe90,CliCon16}   \\
    Sandstones   &   $\geq$15.92   &   $\geq$1.59   &   $\geq$0.16   &     \cite{TelGelShe90,Ves91,KniNur97}   \\
    Bitum. Coal   &   $\geq$12.33   &   $\geq$1.23   &   $\geq$0.12   &     \cite{TelGelShe90,CliCon16,Ves91}   \\
    Anthracite   &   $\geq$0.5   &   $\geq$50.33$\cdot10^{-3}$   &   $\geq$50.33$\cdot10^{-4}$   &     \cite{TelGelShe90}   \\
\hline
\end{tabular}
\end{center}
\caption{Skin depth for different rocks and sediments at three different frequencies: 1~kHz, 100~kHz and 10~MHz (sorted in descending order, by minimum value). \label{table_skin_rocks}}
}
\end{table*}

\begin{table*}[!h]
{\small
\begin{center}
\begin{tabular}{|p{4cm}||c|c|c|c|}
    \hline
{\bf Minerals and mineralized waters} & \multicolumn{3}{|c|}{\bf Skin depth [m]}  & {\bf References} \\
    \cline{2-4}
    ~ & {\bf (f = 1~kHz)} & {\bf (f = 100~kHz)} & {\bf (f = 10~MHz)} & ~ \\
\hline\hline
    Fluorite   &   $\geq$1000   &   $\geq$1000   &   $\geq$1000   &     \cite{TelGelShe90}   \\
    Calcite   &   $\geq$1000   &   $\geq$1000   &   $\geq$1000   &     \cite{TelGelShe90}   \\
    Sylvite   &   $\geq$1000   &   $\geq$1000   &   $\geq$1000   &     \cite{TelGelShe90,LowMar70}   \\
    Quartz   &   $\geq$1000   &   $\geq$1000   &   $\geq$1000   &     \cite{TelGelShe90}   \\
    Petroleum   &   $\geq$1000   &   $\geq$1000   &   $\geq$1000   &     \cite{TelGelShe90,Lin16,KniNur97}   \\
    Ice   &   $\geq$1000   &   $\geq$1000   &   $\geq$1000   &     \cite{TelGelShe90}   \\
    Cinnabar   &   $\geq$1000   &   $\geq$1000   &   $\geq$1000   &     \cite{TelGelShe90}   \\
    Stibnite   &   $\geq$1000   &   $\geq$1000   &   $\geq$1000   &     \cite{TelGelShe90}   \\
    Franklinite   &   $\geq$1000   &   $\geq$1000   &   $\geq$1000   &     \cite{TelGelShe90}   \\
    Limonite   &   $\geq$503.39   &   $\geq$51.31   &   $\geq$14.11   &     \cite{TelGelShe90,RosSmi36}   \\
    Mica   &   $\geq$477.49   &   $\geq$48.05   &   $\geq$8.07   &     \cite{TelGelShe90,CliCon16}   \\
    Bauxite   &   $\geq$225.09   &   22.64--147.2   &   3.75--104.93   &     \cite{TelGelShe90}   \\
    Biotite   &   $\geq$225.08   &   $\geq$22.57   &   $\geq$2.89   &     \cite{TelGelShe90}   \\
    Siderite   &   133.16--133.16   &   13.33--13.33   &   1.52--1.52   &     \cite{TelGelShe90}   \\
    Meteoric waters   &   87.17--503.7   &   8.74--54.58   &   1.1--28.75   &     \cite{TelGelShe90,Vega16}   \\
    Rock salt   &   $\geq$87.17   &   $\geq$8.72   &   $\geq$0.91   &     \cite{TelGelShe90}   \\
    Rutile   &   87.17--355.91   &   8.72--35.88   &   0.92--6.69   &     \cite{TelGelShe90}   \\
    Drinking water   &   71.18--225.18   &   7.15--23.53   &   1.06--9.53   &     \cite{TelGelShe90,Vega16}   \\
    Bismuthinite   &   67.53--380.7   &   6.77--45.76   &   0.81--33.15   &     \cite{TelGelShe90,Mad04}   \\
    Surface waters (sediments)   &   50.33--159.19   &   5.04--16.26   &   0.62--4.69   &     \cite{TelGelShe90}   \\
    Wolframite   &   $\geq$50.33   &   $\geq$5.03   &   $\geq$0.52   &     \cite{TelGelShe90}   \\
    Diamond   &   $\geq$50.33   &   $\geq$5.03   &   $\geq$0.51   &     \cite{TelGelShe90}   \\
    Gypsum/Anhydrite   &   $\geq$50.33   &   $\geq$5.03   &   $\geq$0.51   &     \cite{TelGelShe90,CliCon16,GuiPla_etal10}   \\
    Sphalerite   &   $\geq$19.49   &   $\geq$1.95   &   $\geq$0.2   &     \cite{TelGelShe90}   \\
    Chromite   &   $\geq$15.92   &   $\geq$1.59   &   $\geq$0.16   &     \cite{TelGelShe90}   \\
    Natural waters (sediments)   &   15.92--159.19   &   1.59--16.31   &   0.16--5.01   &     \cite{TelGelShe90}   \\
    Uraninite (pitchblende)   &   15.92--225.08   &   1.59--22.53   &   0.16--2.5   &     \cite{TelGelShe90,KatRab61}   \\
    Natural waters (ign. Rocks)   &   11.25--195   &   1.13--20.22   &   0.11--7.49   &     \cite{TelGelShe90}   \\
    Sea water   &   6.37--15.92   &   0.64--1.59   &   63.67$\cdot10^{-3}$--0.16   &     \cite{TelGelShe90}   \\
    Surface waters (ign. Rocks)   &   5.03--878.15   &   0.5--157.04   &   50.33$\cdot10^{-3}$--149.4   &     \cite{TelGelShe90}   \\
    Saline waters. 3-20\%   &   3.56--6.16   &   0.36--0.62   &   35.6$\cdot10^{-3}$--61.85$\cdot10^{-3}$   &     \cite{TelGelShe90}   \\
    Pyrolusite   &   1.13--50.33   &   0.11--5.04   &   11.26$\cdot10^{-3}$--0.63   &     \cite{TelGelShe90}   \\
    Hematite   &   $\geq$0.92   &   $\geq$91.89$\cdot10^{-3}$   &   $\geq$91.89$\cdot10^{-4}$   &     \cite{TelGelShe90}   \\
    Magnetite   &   0.71--236.65   &   71.18$\cdot10^{-3}$--29.97   &   71.18$\cdot10^{-4}$--23.51   &     \cite{TelGelShe90,Par67}   \\
    Pyrrhotite   &   0.7--3.15   &   70.48$\cdot10^{-3}$--0.32   &   70.48$\cdot10^{-4}$--31.59$\cdot10^{-3}$   &     \cite{TelGelShe90,Par67,RosSmi36}   \\
    Molybdenite   &   $\geq$0.5   &   $\geq$50.33$\cdot10^{-3}$   &   $\geq$50.33$\cdot10^{-4}$   &     \cite{TelGelShe90}   \\
    Cuprite   &   0.5--275.7   &   50.33$\cdot10^{-3}$--27.94   &   50.33$\cdot10^{-4}$--6.52   &     \cite{TelGelShe90}   \\
    Stannite   &   $\geq$0.5   &   50.33$\cdot10^{-3}$--150.33   &   50.33$\cdot10^{-4}$--111.27   &     \cite{TelGelShe90,Ito15}   \\
    Cassiterite   &   $\geq$0.32   &   31.83$\cdot10^{-3}$--271.19   &   31.83$\cdot10^{-4}$--254.6   &     \cite{TelGelShe90}   \\
    Ilmenite   &   0.3--67.26   &   30.08$\cdot10^{-3}$--6.8   &   30.08$\cdot10^{-4}$--1.46   &     \cite{TelGelShe90,RosSmi36}   \\
    Cobaltite   &   0.3--5.03   &   29.78$\cdot10^{-3}$--0.5   &   29.78$\cdot10^{-4}$--50.44$\cdot10^{-3}$   &     \cite{TelGelShe90}   \\
    Graphite   &   0.16--5.74   &   15.92$\cdot10^{-3}$--0.57   &   15.92$\cdot10^{-4}$--57.55$\cdot10^{-3}$   &     \cite{TelGelShe90}   \\
    Titanomagnetite   &   0.13--18.08   &   12.78$\cdot10^{-3}$--1.81   &   12.78$\cdot10^{-4}$--0.19   &     \cite{TelGelShe90,Shu12}   \\
    Chalcocite   &   87.17$\cdot10^{-3}$--12.33   &   87.17$\cdot10^{-4}$--1.23   &   87.17$\cdot10^{-5}$--0.12   &     \cite{TelGelShe90}   \\
    Galena   &   87.17$\cdot10^{-3}$--306.4   &   87.17$\cdot10^{-4}$--33.26   &   87.17$\cdot10^{-5}$--17.69   &     \cite{TelGelShe90}   \\
    Pyrite   &   85.64$\cdot10^{-3}$--275.64   &   85.64$\cdot10^{-4}$--29.47   &   85.64$\cdot10^{-5}$--14.33   &     \cite{TelGelShe90}   \\
    Bornite   &   79.58$\cdot10^{-3}$--11.25   &   79.58$\cdot10^{-4}$--1.13   &   79.58$\cdot10^{-5}$--0.11   &     \cite{TelGelShe90}   \\
    Arsenopyrite   &   71.18$\cdot10^{-3}$--61.64   &   71.18$\cdot10^{-4}$--6.18   &   71.18$\cdot10^{-5}$--0.85   &     \cite{TelGelShe90}   \\
    Chalcopyrite   &   55.13$\cdot10^{-3}$--8.72   &   55.13$\cdot10^{-4}$--0.87   &   55.13$\cdot10^{-5}$--87.76$\cdot10^{-3}$   &     \cite{TelGelShe90}   \\
    Covellite   &   87.17$\cdot10^{-4}$--0.14   &   87.17$\cdot10^{-5}$--14.24$\cdot10^{-3}$   &   87.17$\cdot10^{-6}$--14.24$\cdot10^{-4}$   &     \cite{TelGelShe90,OsoMgb14}   \\
    \hline
\end{tabular}
\end{center}
\caption {Skin depth for different minerals and mineralized waters at three different frequencies: 1~kHz, 100~kHz and 10~MHz (sorted in descending order, by minimum value). \label{table_skin_minerals}}
}
\end{table*}

\subsection{MI Communication Design Guidelines}

\subsubsection{Path-loss of the Communication Link}

\begin{figure*}[h]
\noindent
\centering

\subfloat[Saline Water]{\label{fig::sal_water} \includegraphics[width=0.3\textwidth]{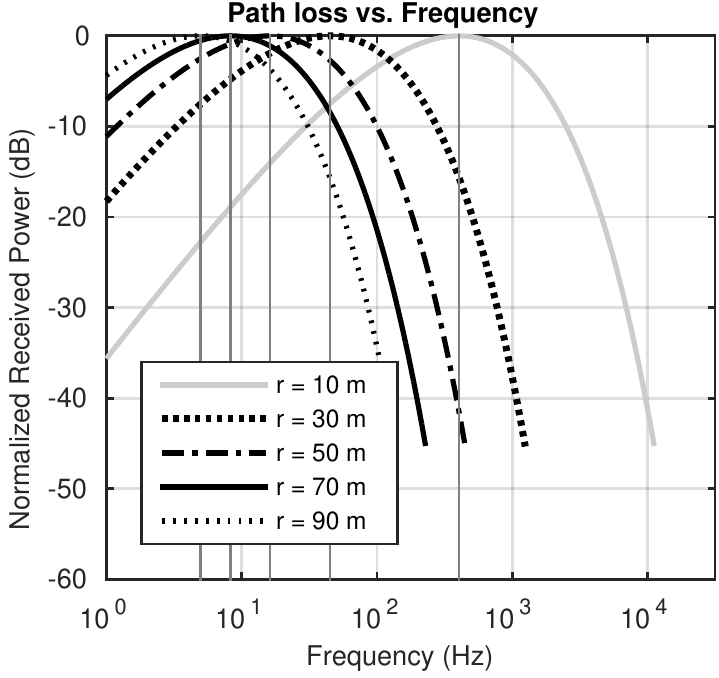}} \hfill
\subfloat[Basalt]{\label{fig::basalt} \includegraphics[width=0.3\textwidth]{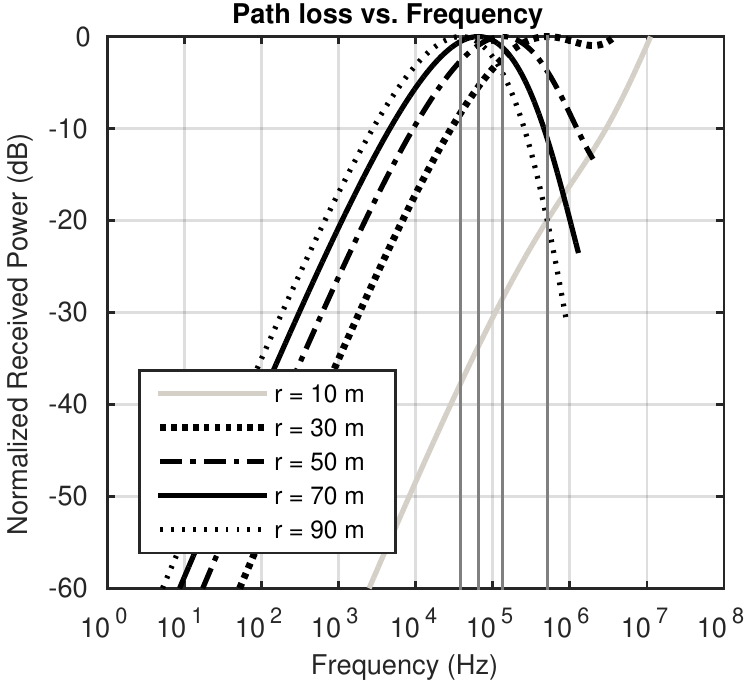}} \hfill
\subfloat[Free Space]{\label{fig::free_space} \includegraphics[width=0.3\textwidth]{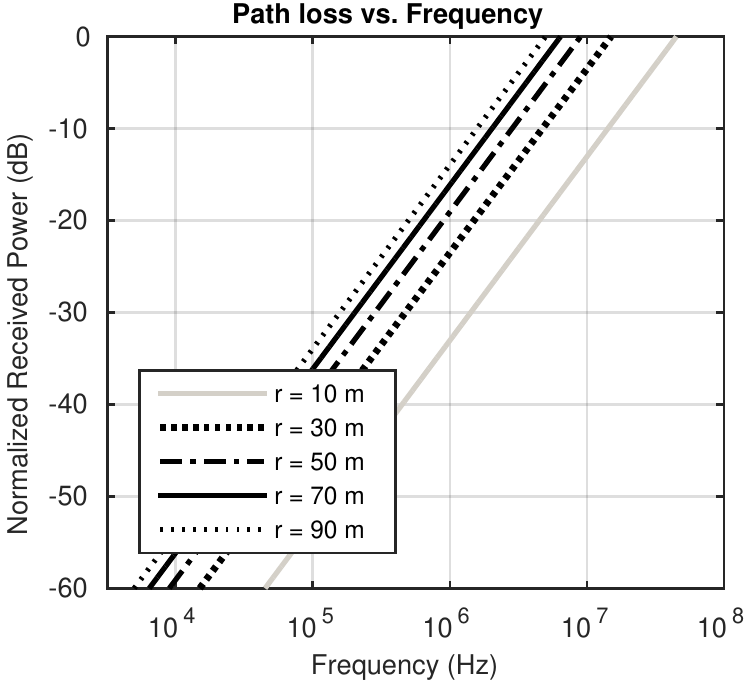}} \\

\caption{Path loss in Eq. \eqref{eqn:attenuation_with_distance} (normalized) as a function of frequency for different separation distances between co-axial TX and RX, each plotted up to $|kr| = 3\pi$ (approximate boundary between near- and far-field), for three different materials, with vertical lines indicating the frequency where maximum received power occurs (optimum frequency). Saline water can be considered a good conductor, which means that for frequencies well within the near-field a bandpass behaviour is seen (caused by the interplay between Faraday's law of induction and the exponential attenuation). Basalt, on the other hand, exhibits a bandpass behaviour for large distances, where the exponential attenuation begins to dominate, and a linear increase of signal with frequency when RX is close to TX. In free space, which is not dissipative, a linear behaviour is seen at all times.}
\label{fig_PathLoss}
\end{figure*}

\begin{figure*}[h]
\noindent
\centering
\subfloat[Optimum Frequency]{\label{fig::opt_freq} \includegraphics[width=0.3\textwidth]{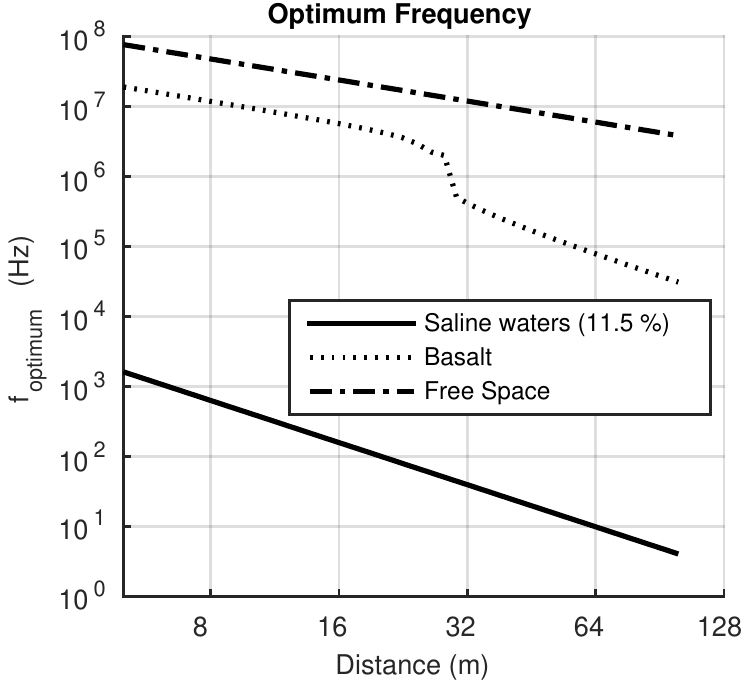}} \hfill
\subfloat[3 dB Bandwidth]{\label{fig::bandwidth} \includegraphics[width=0.3\textwidth]{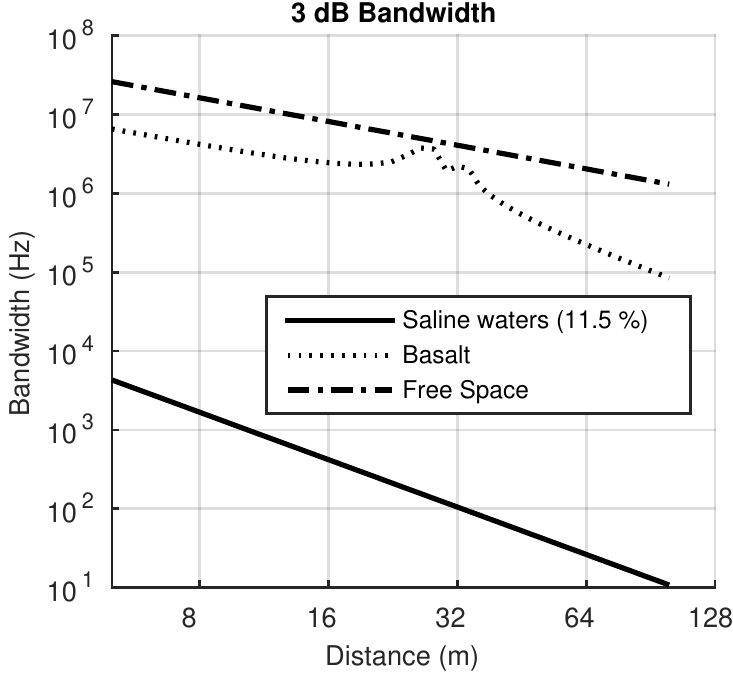}}  \hfill
\subfloat[Path Loss with Optimum Frequency]{\label{fig::pl_opt}\includegraphics[width=0.3\textwidth]{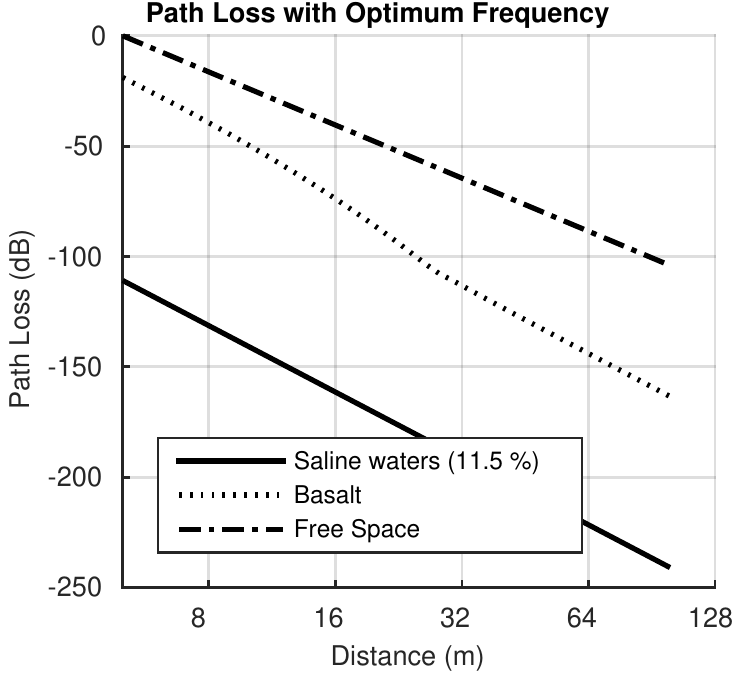}} \\
\caption{Results of optimizing for maximum received power using Eq. \eqref{eqn:attenuation_with_distance}. On the left, optimum frequency is plotted as a function of distance. For all materials, optimum frequency reduces with distance. For small distances the optimum frequency in basalt decays at the same rate for free space, while for larger distances (where the exponential attenuation starts dominating) the rate changes to that of a conductive material, such as saline waters. In the middle, 3 dB bandwidth around optimal frequency is plotted as a function of distance. Again, saline waters exhibit the lowest bandwidth. The peak is due to our definition of 3 dB bandwidth for bandwidth. On the right, the path loss at the optimum frequency is plotted as a function of distance. Saline waters exhibit substantially higher path loss.}
\label{fig_OptimumFrequency}
\end{figure*}

In free-space, the magnitude of the low-frequency field decays more rapidly compared to the high-frequency radio waves (60~dB/decade vs. 20~dB/decade)\cite{Agb11}, 
which calls for a very sensitive RX (or a higher TX power). From the communications perspective, this challenge strengthens the motivation for using multi-hop networks \cite{AbrXiaMarTri16TGRS}. 
However, in other materials, conductivity plays a crucial role, and causes the high-frequency fields to undergo extreme attenuation, much higher than VLF. 
The operation range in different materials is related to the skin depth: for example, at 1~kHz, the skin depth exceeds few tens of meters for the vast majority of rocks and minerals, whereas for higher frequencies (e.g. already at 10~MHz) the skin depth diminishes considerably,
as it will be shown in Section \ref{sec:skin_depth}.

The amplitude of the voltage induced in a coil increases with frequency, as dictated by Faraday's law of induction. 
In free space, where the wavenumber $|k|$ takes on the purely real free space value $k_0$, no dissipation occurs, and the dominant effect in the near field is magnetic induction ($1/r^3$ field decay). 
In materials with finite electrical conductivity, however, the exponential decay term $e^{-k''r}$ attenuates the received flux density $B$, thus reducing the induced voltage. A normalized form of the expression that takes into account both effects is the following:

\begin{equation}
V_{\text{norm}} \sim \frac{2\pi f e^{-k'' r}}{r^3}
\label{eqn:attenuation_with_distance}
\end{equation}

\noindent where $V_{\text{norm}}$ is the magnitude of the induced voltage normalized to the peak value (this expression is valid for co-axial alignment of loops~\cite{SunAky10}, and for triaxial coils~\cite{AbrXiaMarTri16TGRS}). Fig. \ref{fig_PathLoss} shows the attenuation of TX power in dB as a function of frequency, for different distances, for a good conductor (saline water), worse conductor (basalt), and insulator (free space), up to the boundary between near- and far-field (where $|kr| = 2\pi$). A band-pass characteristic is evident in the presence of electrically conductive materials (saline waters); as the conductivity is decreased (basalt), exponential attenuation of the flux density subsides and is overshadowed by the increase of the received voltage due to induction.

\subsubsection{Optimal Communication Frequency and Bandwidth}

One important aspect to point out is that, for conductive materials, the optimal frequency (corresponding to the minimum attenuation) not only depends on the electromagnetic properties of the material, but also on the distance between TX and RX coils, and their relative orientation.
This makes the design of underground wireless sensor networks~\cite{AkyStu06} very challenging~\cite{TanSunAky15,KisAkyGer14}.
Conversely, for a given frequency, there is an optimal communication distance $r_\star$, for which the attenuation is minimized. 
Gibson \cite{Gib10} showed that for good conductors, the optimal distance between TX and RX is in order of few skin depths $\delta$,
i.e., 
\begin{equation} \label{opt_dist}
r_\star=T\delta,
\end{equation} 
where $T$ is the number of skin depths. 
For co-axially aligned coils (and also for triaxial coils~\cite{AbrXiaMarTri16TGRS}), the optimal distance is $r_\star=2.83\delta$  (i.e., $T=2.83$), which corresponds to $r_\star=0.45\lambda$. 
For co-planar coils, $r_\star=3.86\delta$  (i.e., $T=3.86$), which corresponds to $r_\star=0.61\lambda$.
In order to avoid the alignment issue, triaxial coils~\cite{AbrXiaMarTri16TGRS} may be used, hence the voltage becomes orientation invariant (which is equivalent to always co-axially aligned coils). 
Given the operation frequency, the communication distance in certain environments can be roughly predicted as a function of skin depth from Figures~\ref{fig_skin_rocks} and~\ref{fig_skin_minerals}.
Solving for the optimal frequency $f$, given a distance $r$, yields:
\begin{equation} \label{opt_freq}
f = \frac{T^2}{\pi r^2 \sigma \mu }
\end{equation}
However, the above expression is only valid for good conductors, and for distances up to the transition zone ($r <  \lambda$). 
As conductivity is decreased, the effect of the eddy currents becomes less significant. 
Consequently, the local maximum begins to disappear, and the optimum frequency becomes difficult to pinpoint. 
We define the optimum frequency as the frequency that falls halfway between the considered boundary (in our case we chose $|kr| = 2\pi$) and the $-3$~dB cutoff frequency. Increasing the frequency far beyond the limit $|kr| = 2\pi$ will invalidate the near-field magnetic dipole equation, as radiation will occur.

Fig. \ref{fig::opt_freq} shows the variation of optimal frequency with distance, and Fig. \ref{fig::bandwidth} shows the path loss at the optimal frequency. Fig \ref{fig::pl_opt} shows the attenuation w.r.t. distance at the corresponding optimal frequencies. These plots illustrate how materials behave differently at different frequencies, which will also influence the choice of optimal frequency considering the distance between nodes.

\subsection{MI Localization Design Guidelines}
For MI localization systems, the accuracy of quasi-static approximation is crucial, since the source can be approximated by a point magnetic dipole. Therefore, the most important parameter is the operation frequency.
In conductive media, the operation frequency should be as low as possible in order to avoid eddy currents that produce strong secondary fields, thus invalidating the magnetic dipole equations, but sufficiently high to achieve the desired bandwidth (i.e., provide location estimates at sufficient rate).
Gibson~\cite{Gib10} showed that the quasi-static model may be successfully applied for localization using audio range frequencies at distances that are smaller than one third of a skin depth, i.e. 
\begin{equation} \label{r_loc}
r < \delta/3.
\end{equation}
Condition \eqref{r_loc} minimizes that the distortion of the vector fields components, and therefore ensures reasonable localization accuracy using the magnetic dipole equations. 
However, given a low operation frequency, this condition is in conflict with the optimal distance (that ensures minimum attenuation), which is few skin depths, as shown by Eq.~\eqref{opt_dist}.
Therefore, the system is constrained to operate below the optimal frequency in Eq.~\eqref{opt_freq}, in the lower part of the spectrum, where attenuation is higher due to the Faraday law. Consequently, the operation range will be decreased, imposing either higher transmit power, or higher RX sensitivity. 
Given an operation frequency, Tables~\ref{table_skin_rocks} and~\ref{table_skin_minerals} may be used to determine the achievable operation range of a localization system in different environments as $\delta/3$.
Figures~\ref{fig_skin_rocks} and~\ref{fig_skin_minerals} may be used for a quicker view, and it includes not only the minimum and maximum skin depth values, but also the average value for most materials. 


\section{Related work}
MI communication~\cite{GabDegWai71,Agb11} is a promising technology that can operate reliably in challenging environments that are practically inaccessible to radio waves, such as underground and underwater. 
Having a good channel model is crucial in the design phase of the MI system, in order to ensure reliable operation. 
MI channel sounding ~\cite{Gib10} is typically done either by using tuned resonant circuits (narrowband), or untuned circuits (wideband).
In order to achieve high energy efficiency, resonant coupling typically involves a high Q-factor (quality factor). 
Consequently, the resulting channel exhibits a much narrower bandwidth compared to the transmission medium alone. 
However, we are interested in theoretically characterizing the wideband channel frequency response of the medium.
Wideband channel sounding using narrowband resonant coils has been proposed, and involves frequency stepping~\cite{Gib10}. 
Wideband untuned transceivers are also being used, but they are energy inefficient. 
Consequenctly, the corresponding SNR is very low, which calls for pseudorandom codes and averaging over long periods of time~\cite{Gib10}. 
Most literature~\cite{SunAky10,KisAkyGer14,LinAkyWanSun15,TanSunAky15} addresses the channel modeling from an end-to-end
perpective, i.e. the channel that contains not only the transmission medium, but also the TX and RX coils. 
In this paper, we focus strictly on the medium-related issues, leaving the transceiver and coil design up to the system designer, thus offering more design flexibility. 
Separating the TX/RX coil circuits from the medium has also been considered in~\cite{Gib10,SilMog15}.

The work in \cite{SunAky10} addresses the problem of steep decay of the MI field magnitude in MI communication by using a waveguide comprised of aligned mono-axial passive resonators. 
The approach in~\cite{KisAkyGer14} aims to maximize the network throughput of a relay system that involves a cascade of passive resonators (as in~\cite{SunAky10}) subject to carrier frequency, coil number of turns, number of links, as well as to reduce the interference by finding optimal coil orientations. The problem is formulated as a multivariate optimization. 
In \cite{LinAkyWanSun15}, an environment-aware cross-layer system design is proposed, whose final goal is to maximize the quality of service (packet delay and transmission reliability), which is achieved by optimizing a composite cost function that includes throughput and energy consumption. Direct sequence code division multiple access with distributed power control that relies on a non-cooperative game is employed. Geographic routing is used to forward the packets. Although the scheme is designed to be environment-aware, it does not select the operation frequency according to the electro-magnetic properties of the transmission medium. The communication frequency is fixed to 7~MHz for all nodes, being considered suitable for underground communication (according to~\cite{SunAky10}) and therefore, it is not included in the cross-layer optimization. The optimum operation frequency expression we are providing in this paper may be used to furhter adapt the system to the environment. 

In \cite{SilMog15}, the optimization of the operation frequency of underground wireless sensor networks is addressed, but only soil medium is considered. 
In addition, the optimal frequency is not provided in closed-form, but rather determined by a grid search, given various design parameters. 
RX voltage values for differently moisturized soils are tabulated considering particular values of the coil parameters. 
It is pointed out that audio frequencies are suitable for mid-range communication (15 to 30 meters) in conductive media such as high soil moisture. 
In addition, it is proposed that the TX/RX coils might require two different wirings in order to adapt to different soil moisture conditions. 
An adaptive frequency MI network using axially aligned coils is proposed in~\cite{SilMog15a}. 
Inspired by~\cite{Gib10} and \cite{SilMog15}, in this paper we provide closed-form expression of the optimal frequency that can be used in various underground environments.
We also provide skin depth values for various rocks and minerals. 

ITU-R Standard~\cite{ITU-R_R527} includes skin depth values for sea water with different salinity, fresh water, water ice and dry to moist soils for the frequency range 100~kHz--100~GHz, 
but no rocks and minerals are considered.
A similar, but more comprehensive collection of attenuation values for the same materials as in~\cite{ITU-R_R527} is provided in~\cite{ITU-R_R368}, along with
generic attenuation values for different combinations of electric conductivity, permittivity and magnetic permeability values, for frequency range 10~kHz--30~MHz.
This is not particularly useful for the magneto-inductive localization or communications research communities, since one must look up for the right combination of electromagnetic constants 
that fits the materials of interest.
Moreover, the assumptions made in\cite{ITU-R_R368} are not appropriate for the underground magneto-inductive communications and localization applications we are considering in this paper, for several reasons:
1) an electric antenna (vertical monopole) is assumed; 
2) transmitter and receiver are at the ground level; 
3) the curves are provided for the vertical field strength component of the radiated field (i.e. the far field region)
By contrast, we consider induction loop antennas both at transmitter and receiver, and operation in the near-field region, through the ground.
In addition, in order to predict how certain material behaves at a given frequency, there is no need to look up for electromagnetic constants in tables.

Finally, we also provide conditions for reliable operation of a MI localization system that relies on the magnetic dipole equations, as a function of skin depth. 
In~\cite{DavCheTucAtk08}, it was shown that accurate localization can be achieved even beyond one skin depth, by exploiting the geometric properties of the magnetic fields, and assuming horizontal underground layers.
More sophisticated models for conductive media have also been proposed in the literature, such as 
homogeneous earth model~\cite{SogVicVan_etal04,DavCheTucAtk08}, 
stratified earth model~\cite{Wai71}, 
and image theory model~\cite{KypAbrMar15Sens} in highly distorted environments.
The comprehensive collection of skin depth values for the most common rocks and minerals provided in this paper is beneficial in the design of MI communications and localization systems that operate under the ground, as well as to predict the reliability of such system given its operation parameters.


\section{Discussion and Conclusions}
In this paper, we provide attenuation figures for most common underground materials.
In addition, we provide the tabulated raw values of electromagnetic constants, and the source code used in this paper~\cite{Mathworks_underground}, in order to help
{\em MI communications} and {\em MI localization} research communities to easily obtain basic system design guidelines. 
For MI communication, we provide optimal operation frequencies given the distance between nodes, and the achievable bandwidth at those frequencies. 
We also provide the attainable operation range for a MI localization system that relies on dipole equations.
Tables \ref{table_skin_rocks} and \ref{table_skin_minerals} show that at very-low frequencies, the skin effect is negligible in most underground materials. 
The simple magnetic dipole model can be still used in most common underground scenarios, since we are still operating within the near field region~\cite{DavCheTucAtk08,SogVicVan_etal04}.
The tabulated skin depth values may be directly used to predict the behavior of a MI system in certain environments, i.e.,
roughly estimate the achievable range, operation frequencies, bandwidth, path loss, and distortions of the vector field components in conductive media.

\bibliographystyle{IEEEtran} 
{\small


\begin{thebibliography}{10}
\providecommand{\url}[1]{#1}
\csname url@rmstyle\endcsname
\providecommand{\newblock}{\relax}
\providecommand{\bibinfo}[2]{#2}
\providecommand\BIBentrySTDinterwordspacing{\spaceskip=0pt\relax}
\providecommand\BIBentryALTinterwordstretchfactor{4}
\providecommand\BIBentryALTinterwordspacing{\spaceskip=\fontdimen2\font plus
\BIBentryALTinterwordstretchfactor\fontdimen3\font minus
  \fontdimen4\font\relax}
\providecommand\BIBforeignlanguage[2]{{%
\expandafter\ifx\csname l@#1\endcsname\relax
\typeout{** WARNING: IEEEtran.bst: No hyphenation pattern has been}%
\typeout{** loaded for the language `#1'. Using the pattern for}%
\typeout{** the default language instead.}%
\else
\language=\csname l@#1\endcsname
\fi
#2}}

\bibitem{GabDegWai71}
R.~Gabillard, P.~Degauque, and J.~Wait, ``Subsurface electromagnetic
  telecommunication--a review,'' \emph{IEEE Transactions on Communication
  Technology}, vol.~19, no.~6, pp. 1217--1228, December 1971.

\bibitem{SunAky10}
Z.~Sun and I.~Akyildiz, ``Magnetic induction communications for wireless
  underground sensor networks,'' \emph{Antennas and Propagation, IEEE
  Transactions on}, vol.~58, no.~7, pp. 2426--2435, July 2010.

\bibitem{SojWraDin01}
J.~J. Sojdehei, P.~N. Wrathall, and D.~F. Dinn, ``{Magneto-inductive (MI)
  communications},'' in \emph{Oceans 2001, MTS/IEEE Conference, Honolulu, USA},
  2001.

\bibitem{MarTri12}
A.~Markham and N.~Trigoni, ``Magneto-inductive networked rescue system
  {(MINERS)}: taking sensor networks underground,'' in \emph{"11th
  International Conference on Information Processing in Sensor Networks (IPSN
  2012)}, 2012.

\bibitem{KisAkyGer14}
S.~Kisseleff, I.~F. Akyildiz, and W.~H. Gerstacker, ``Throughput of the
  magnetic induction based wireless underground sensor networks: Key
  optimization techniques,'' \emph{IEEE Transactions on Communications},
  vol.~62, no.~12, pp. 4426--4439, Dec. 2014.

\bibitem{AkySunVur09}
I.~F. Akyildiz, Z.~Sun, and M.~C. Vuran, ``Signal propagation techniques for
  wireless underground communication networks,'' \emph{Physical Communication},
  vol.~2, no.~3, 2009.

\bibitem{TanSunAky15}
X.~Tan, {Zhi Sun}, and I.~F. Akyldiz, ``Wireless underground sensor networks,''
  \emph{IEEE Antennas and Propagation Magazine}, vol.~57, no.~4, pp. 74--87,
  Aug. 2015.

\bibitem{AbrXiaMarTri16TGRS}
T.~E. Abrudan, Z.~Xiao, A.~Markham, and N.~Trigoni, ``Underground,
  incrementally deployed magneto-inductive {3-D} positioning network,''
  \emph{IEEE Transactions on Geoscience and Remote Sensing}, pp. 1--16, 2016,
  (to appear).

\bibitem{AbrXiaMarTri15JSAC}
------, ``Distortion rejecting magneto-inductive three-dimensional localization
  ({MagLoc}),'' \emph{IEEE Journal on Selected Areas in Communications},
  vol.~33, no.~11, pp. 2404--2417, Nov. 2015.

\bibitem{MarTriEllMac12}
A.~Markham, N.~Trigoni, D.~W. Macdonald, and S.~A. Ellwood, ``Underground
  localization in {3-D} using magneto-inductive tracking,'' \emph{IEEE Sensors
  Journal}, vol.~12, no.~6, pp. 1809--1816, June 2012.

\bibitem{MarTriEllMac10}
A.~Markham, N.~Trigoni, S.~A. Ellwood, and D.~W. Macdonald, ``Revealing the
  hidden lives of underground animals using magneto-inductive tracking,'' in
  \emph{8th ACM Conference on Embedded Networked Sensor Systems (Sensys 2010)},
  Z\"urich, Switzerland, Nov. 2010.

\bibitem{DavCheTucAtk08}
C.~Davis, W.~C. Chew, W.~Tucker, and P.~Atkins, ``A null-field method for
  estimating underground position,'' \emph{IEEE Transactions on Geoscience and
  Remote Sensing}, vol.~46, no.~11, pp. 3731--3738, Nov 2008.

\bibitem{SogVicVan_etal04}
J.~Sogade, Y.~Vichabian, A.~Vandiver, P.~Reppert, D.~Coles, and F.~Morgan,
  ``Electromagnetic cave-to-surface mapping system,'' \emph{IEEE Transactions
  on Geoscience and Remote Sensing}, vol.~42, no.~4, pp. 754--763, April 2004.

\bibitem{AkyStu06}
I.~F. Akyildiz and E.~P. Stuntebeck, ``Wireless underground sensor networks:
  Research challenges,'' \emph{Ad Hoc Net.}, vol.~4, no.~6, 2006.

\bibitem{Gib10}
A.~D.~W. Gibson, ``Channel characterisation and system design for sub-surface
  communications,'' Ph.D. dissertation, School of Electronic and Electrical
  Engineering, University of Leeds, May 2010, (revised Feb. 2004 version).

\bibitem{LinAkyWanSun15}
S.~Lin, I.~F. Akyldiz, {Pu Wang}, and {Zhi Sun}, ``Distributed cross-layer
  protocol design for magnetic induction communication in wireless underground
  sensor networks,'' \emph{IEEE Transactions on Wireless Communications},
  vol.~14, no.~7, pp. 4006--4019, Jul. 2015.

\bibitem{PriHow04}
E.~Prigge and J.~How, ``Signal architecture for a distributed magnetic local
  positioning system,'' \emph{IEEE Sensors Journal}, vol.~4, no.~6, pp.
  864--873, Dec. 2004.

\bibitem{LiVurAky07}
L.~Li, M.~C. Vuran, and I.~F. Akyildiz, ``Characteristics of underground
  channel for wireless underground sensor networks,'' in \emph{6th Annual
  Mediterranean Ad Hoc Networking Workshop}, Corfu, Greece, 2007.

\bibitem{AyuCucLerVil10}
N.~Ayuso, J.~Cuchi, F.~Lera, and J.~Villarroel, ``Accurately locating a
  vertical magnetic dipole buried in a conducting earth,'' \emph{Geoscience and
  Remote Sensing, IEEE Transactions on}, vol.~48, no.~10, pp. 3676--3685, Oct
  2010.

\bibitem{Mathworks_underground}
T.~E. Abrudan, ``Impact of rocks and minerals on underground magneto-inductive
  communication and localization, {Matlab} source code and tables of
  electromagnetic constants,'' Mathworks {MATLAB} Central File Exchange, Apr.
  2016, [Online]:
  https://www.mathworks.com/matlabcentral/fileexchange/56632-impact-of-rocks-and-minerals-on-underground-magneto-inductive-communication-and-localization.

\bibitem{Wai71}
J.~R. Wait, \emph{Electromagnetic Waves in Stratified Media}.\hskip 1em plus
  0.5em minus 0.4em\relax New York: Pergamon, 1971.

\bibitem{ITU-R_R527}
{International~Telecommunication~Union~(ITU-T)}, ``Electrical characteristics
  of the sufrace of earth,'' {Recommendation ITU-R~P.527-3}, Mar. 1992.

\bibitem{Bla96}
R.~J. Blakely, \emph{Potential Theory in Gravity and Magnetic
  Applications}.\hskip 1em plus 0.5em minus 0.4em\relax Cambridge University
  Press, 1996.

\bibitem{TelGelShe90}
W.~M. Telford, L.~P. Geldart, and R.~E. Sheriff, \emph{Applied Geophysics (2nd
  edition)}.\hskip 1em plus 0.5em minus 0.4em\relax Cambridge University Press,
  1990.

\bibitem{JorBal68}
E.~C. Jordan and K.~G. Balmain, \emph{Electromagnetic Waves and Radiating
  Systems}.\hskip 1em plus 0.5em minus 0.4em\relax Prentice-Hall, Englewood
  Cliffs, NJ, 1968, vol.~4.

\bibitem{Agb11}
J.~I. Agbinya, \emph{Principles of Inductive Near Field Communications for
  Internet of Things}, ser. River Publishers Series on Communications.\hskip
  1em plus 0.5em minus 0.4em\relax River Publishers, 2011.

\bibitem{Par67}
E.~I. Parkhomenko, \emph{Electrical properties of rocks}, ser. Monographs in
  Science, G.~V. Keller, Ed.\hskip 1em plus 0.5em minus 0.4em\relax Springer,
  1967.

\bibitem{Cla66}
S.~P. {Clark~Jr.}, \emph{Handbook of Physical Constants (Revised
  Edition)}.\hskip 1em plus 0.5em minus 0.4em\relax Yale University, New Haven,
  Connecticut: The Geological Society of America (Memoir 97), 1966.

\bibitem{BirSchSpi50}
F.~Birch, J.~F. Schairer, and H.~C. Spicer, Eds., \emph{Handbook of Physical
  Constants}.\hskip 1em plus 0.5em minus 0.4em\relax Geological Society of
  America, 1950.

\bibitem{NRC_etal_27}
{National Research Council (U.S.)}, {International Council of Scientific
  Unions}, and {National Academy of Sciences (U.S.)}, \emph{{International
  Critical Tables of Numerical Data, Physics, Chemistry and Technology}}, E.~W.
  Washburn, Ed.\hskip 1em plus 0.5em minus 0.4em\relax National Academies,
  1927, vol.~2.

\bibitem{RosSmi36}
J.~L. Rosenholtz and D.~T. Smith, ``The dielectric constants of mineral
  powders,'' \emph{The American Mineralogist}, vol.~21, no.~2, 1936.

\bibitem{Shu12}
R.~Shuey, \emph{Semiconducting Ore Minerals}, ser. Developments in Economic
  Geology.\hskip 1em plus 0.5em minus 0.4em\relax Elsevier, 2012, vol.~4.

\bibitem{Pat15}
{Imke~de~Pater} and J.~J. Lissauer, \emph{Planetary Sciences}, 2nd~ed.\hskip
  1em plus 0.5em minus 0.4em\relax Cambridge University Press, 2015.

\bibitem{Spe05}
J.~G. Speight, \emph{Handbook of Coal Analysis}, ser. Chemical Analysis: A
  Series of Monographs on Analytical Chemistry and Its Applications.\hskip 1em
  plus 0.5em minus 0.4em\relax John Wiley \& Sons, 2005, vol. 166.

\bibitem{Ves91}
V.~A. Veksler and A.~V. Bashirov, ``Statistical models of the probability
  distribution of the dielectric parameters of rock,'' \emph{Journal of Mining
  Science}, vol.~27, no.~1, pp. 62--66, 1991.

\bibitem{KniNur97}
R.~J. Knight and A.~Nur, ``The dielectric constant of sandstones, {60~kHz} to
  {4~MHz},'' \emph{Geophysics}, vol.~52, no.~5, pp. 644--654, 1987.

\bibitem{AhmZih90}
M.~S. Ahmad and A.~M. Zihlif, ``Some magnetic and electrical properties of
  basalt rocks,'' \emph{Materials Letters}, vol.~10, no. 4--5, pp. 207--214,
  Nov. 1990.

\bibitem{Gai14}
R.~Gaikwad, ``Electrical properties estimation of oil sands,'' Master's thesis,
  Department of Chemical and Materials Engineering, University of Alberta,
  2014.

\bibitem{GuiPla_etal10}
A.~Guinea, L.~R. E.~Play\`a, M.~Himi, and R.~Bosch, ``Geoelectrical
  classification of gypsum rocks,'' \emph{Surveys in Geophysics}, vol.~31,
  no.~6, pp. 557--580, 2010.

\bibitem{Ito15}
K.~Ito, \emph{Copper Zinc Tin Sulfide-Based Thin Film Solar Cells}.\hskip 1em
  plus 0.5em minus 0.4em\relax John Wiley \& Sons, 2015.

\bibitem{LowMar70}
R.~P. Lowndes and D.~H. Martin, ``Dielectric constants of ionic crystals and
  their variations with temperature and pressure,'' in \emph{Proceedings of the
  Royal Society of London}, ser. A, Mathematical and Physical, vol. 316, no.
  1526, May 1970, pp. 351--375.

\bibitem{KatRab61}
J.~J. Katz and E.~Rabinowitch, \emph{The Chemistry of Uranium}.\hskip 1em plus
  0.5em minus 0.4em\relax Ripol Classic, 1961.

\bibitem{Mad04}
O.~Madelung, \emph{Semiconductors Data Handbook}.\hskip 1em plus 0.5em minus
  0.4em\relax Springer, Berlin, 2004.

\bibitem{OsoMgb14}
J.~C. Osuwa and E.~C. Mgbaja, ``Structural and electrical properties of copper
  sulfide {(CuS)} thin films doped with mercury and nickel impurities,''
  \emph{IOSR Journal of Applied Physics}, vol.~6, p.~5, 2014.

\bibitem{Mar_12}
\emph{The Marconigraph}.\hskip 1em plus 0.5em minus 0.4em\relax {Marconi's
  Wireless Telegraph Company}, 1912.

\bibitem{Lin16}
M.~Lindner, ``Oil condition monitoring using electrical conductivity,''
  Machinery Lubrication, URL:
  http://www.machinerylubrication.com/Read/29407/oil-condition-monitoring,
  Retrieved, Mar. 2016.

\bibitem{CliCon16}
{Clipper~Controls}, ``Dielectric constant values,'' URL:
  http://www.clippercontrols.com/pages/Dielectric-Constant-Values.html,
  Retrieved, Mar. 2016.

\bibitem{Vega16}
{Vega~Americas~Inc.}, ``Dielectric constants list,'' {URL:
  http://www.vega-americas.com/downloads/Forms-Certificates/Dielectric\_Constants\_List.pdf},
  Retrieved, Mar. 2016.

\bibitem{SilMog15}
A.~R. Silva and M.~Moghaddam, ``Operating frequency selection for low-power
  magnetic induction-based wireless underground sensor networks,'' in
  \emph{Sensors Applications Symposium (SAS), 2015 IEEE}, April 2015, pp. 1--6.

\bibitem{SilMog15a}
------, ``Strategic frequency adaptation for mid-range magnetic induction-based
  wireless underground sensor networks,'' in \emph{Systems Conference (SysCon),
  2015 9th Annual IEEE International}, April 2015, pp. 758--765.

\bibitem{ITU-R_R368}
{International~Telecommunication~Union~(ITU-T)}, ``Ground-wave propagation
  curves for frequencies between {10~kHz} and {30~MHz},'' {Recommendation
  ITU-R~P.368-9}, Feb. 2007.

\bibitem{KypAbrMar15Sens}
O.~Kypris, T.~E. Abrudan, and A.~Markham, ``Reducing magneto-inductive
  positioning errors in a metal-rich indoor environment,'' in \emph{IEEE
  Sensors Conference}, 1--4 Nov. 2015, pp. 1--4.

\end{thebibliography}
}


\begin{IEEEbiography}[{\includegraphics[width=1in,height=1.25in,clip,keepaspectratio]{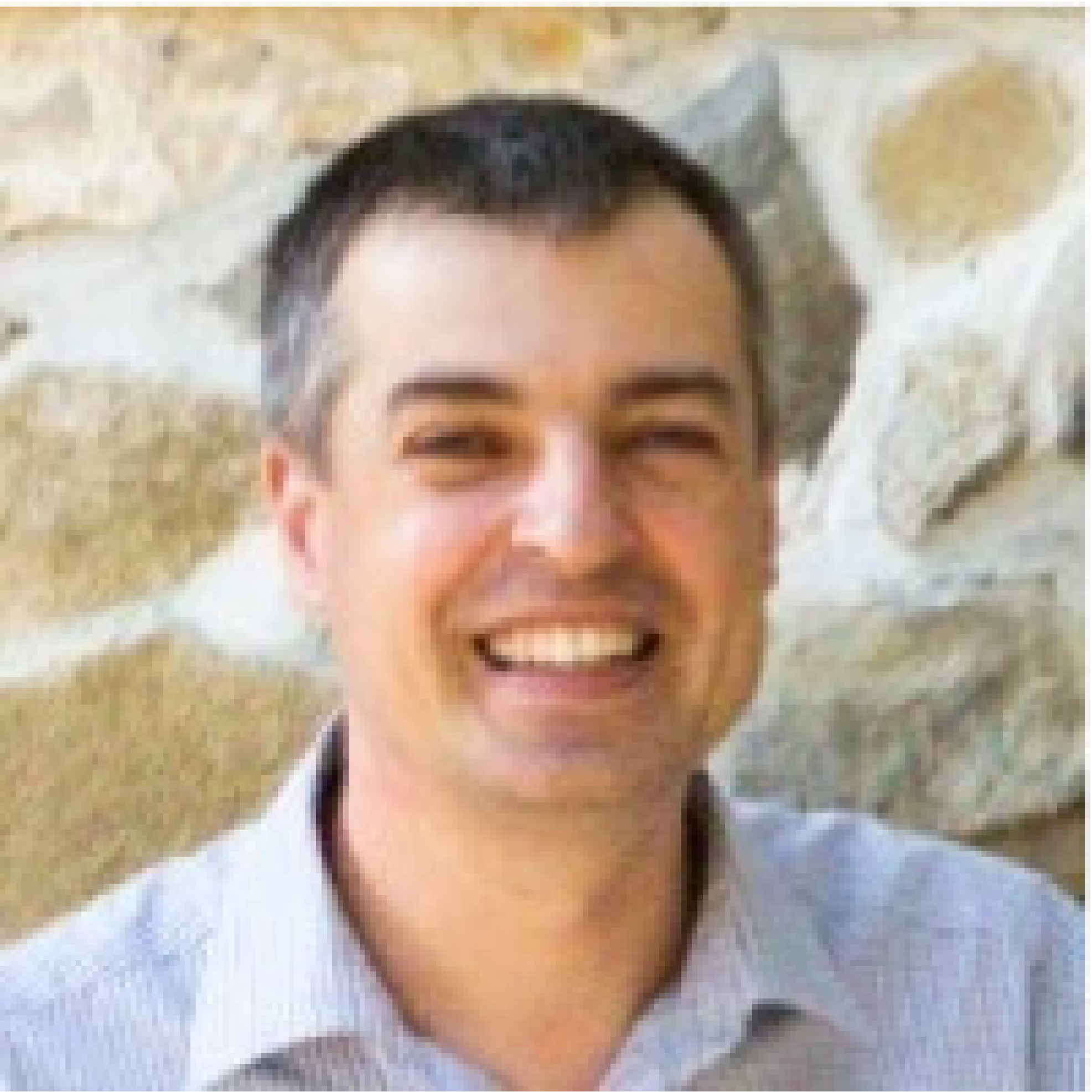}}]
{Traian E. Abrudan} (S'02--M'09) received the D.Sc. degree (with honors) from Aalto University, Finland (2008), and the M.Sc. degree from the Technical University of Cluj-Napoca, Romania (2000). 
During 2010--2013, he was a postdoctoral researcher at the Faculty of Engineering, University of Porto, and a member of Instituto de Telecomunica\c{c}\~oes, Portugal.
Since 2013, he has been a postdoctoral researcher at the Department of Computer Science, University of Oxford, 
working on practical localization algorithms and systems for humans and robots using low-frequency magnetic fields, as well as other sensing modalities. 
His fundamental research topics include sensor array signal processing, applied parameter estimation, numerical optimization, and wireless transceiver algorithms.
\end{IEEEbiography}

\begin{IEEEbiography}[{\includegraphics[width=1in,height=1.25in,clip,keepaspectratio]{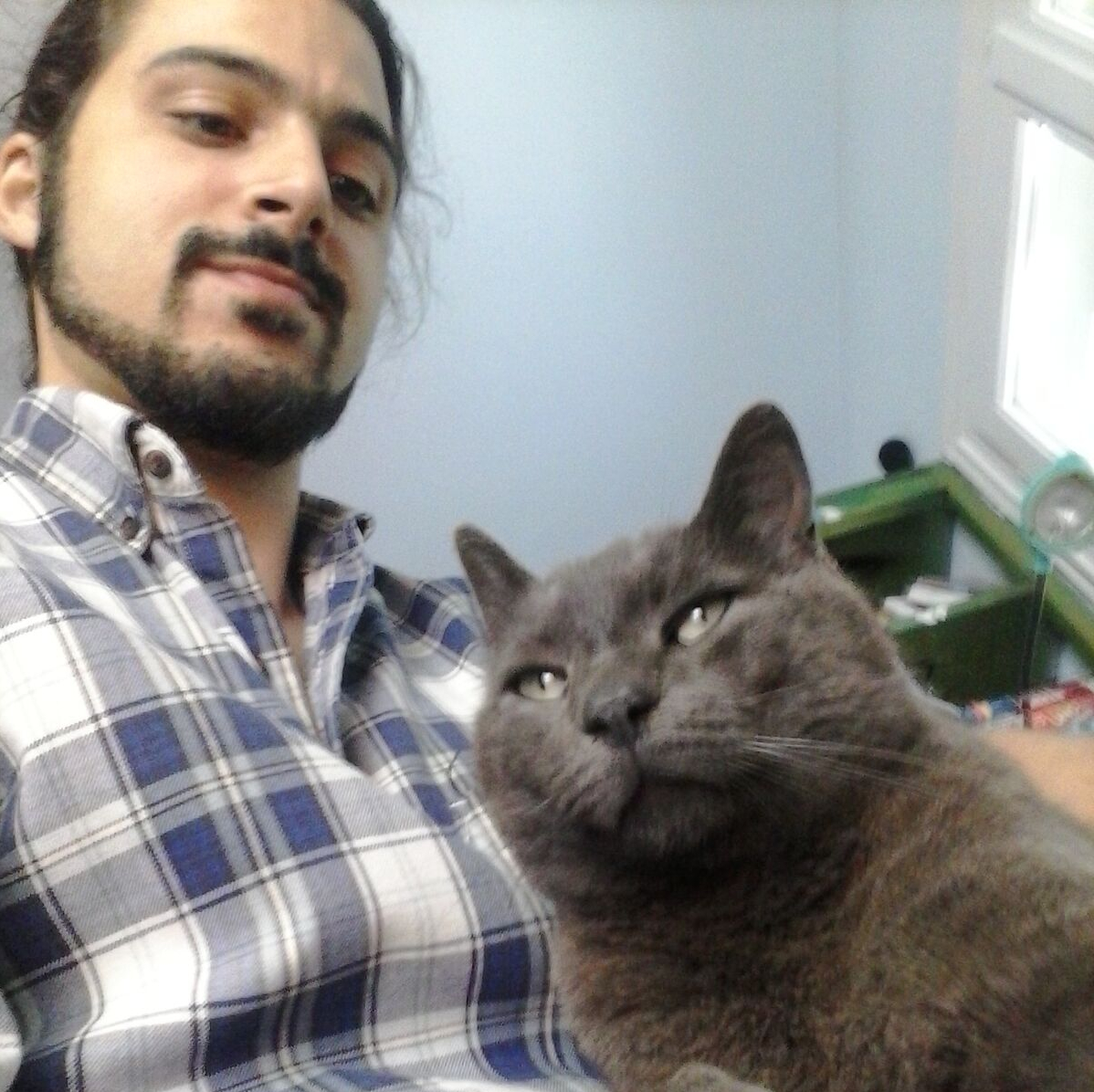}}]
{Orfeas Kypris} (S'11--M'15) received the BEng degree in Electrical and Electronic Engineering in 2009, and the MSc degree in Magnetics in 2010 from Cardiff University, Cardiff, U.K. He then joined the Department of Electrical and Computer Engineering at Iowa State University, U.S.A., where he obtained his Ph.D. in Electrical Engineering in 2015. Since 2015, he has been a postdoctoral researcher at the Department of Computer Science, University of Oxford, working on indoor localization and structural health monitoring using low-frequency magnetic fields. His research interests include non-destructive evaluation using Barkhausen signals, applied electromagnetism and magnetic materials. He is a member of the IEEE Eta Kappa Nu (IEEE HKN), and the IEEE Magnetics Society.
\end{IEEEbiography}

\begin{IEEEbiography}[{\includegraphics[width=1in,height=1.25in,clip,keepaspectratio]{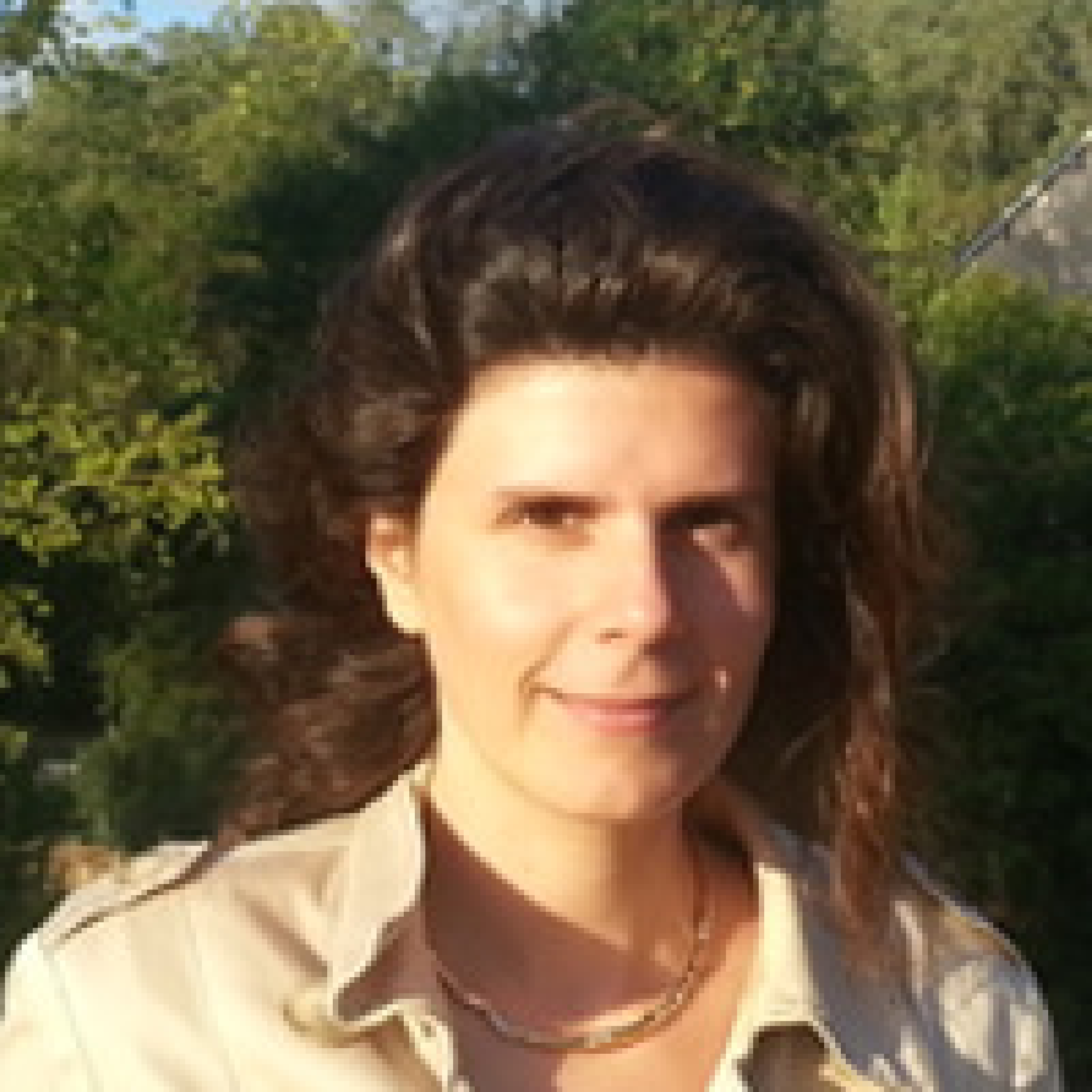}}]
{Dr. Niki Trigoni} is a Professor at the Department of Computer Science, University of Oxford. She obtained her PhD at the University of Cambridge (2001), became a postdoctoral researcher at Cornell University (2002-2004), and a Lecturer at Birkbeck College (2004-2007). Since she moved to Oxford in 2007, she established the Sensor Networks Group, and has conducted research in communication, localization and in-network processing algorithms for sensor networks. Her recent and ongoing projects span a wide variety of sensor networks applications, including indoor/underground localization, wildlife sensing, road traffic monitoring, autonomous (aerial and ground) vehicles, and sensor networks for industrial processes.
\end{IEEEbiography}

\begin{IEEEbiography}[{\includegraphics[width=1in,height=1.25in,clip,keepaspectratio]{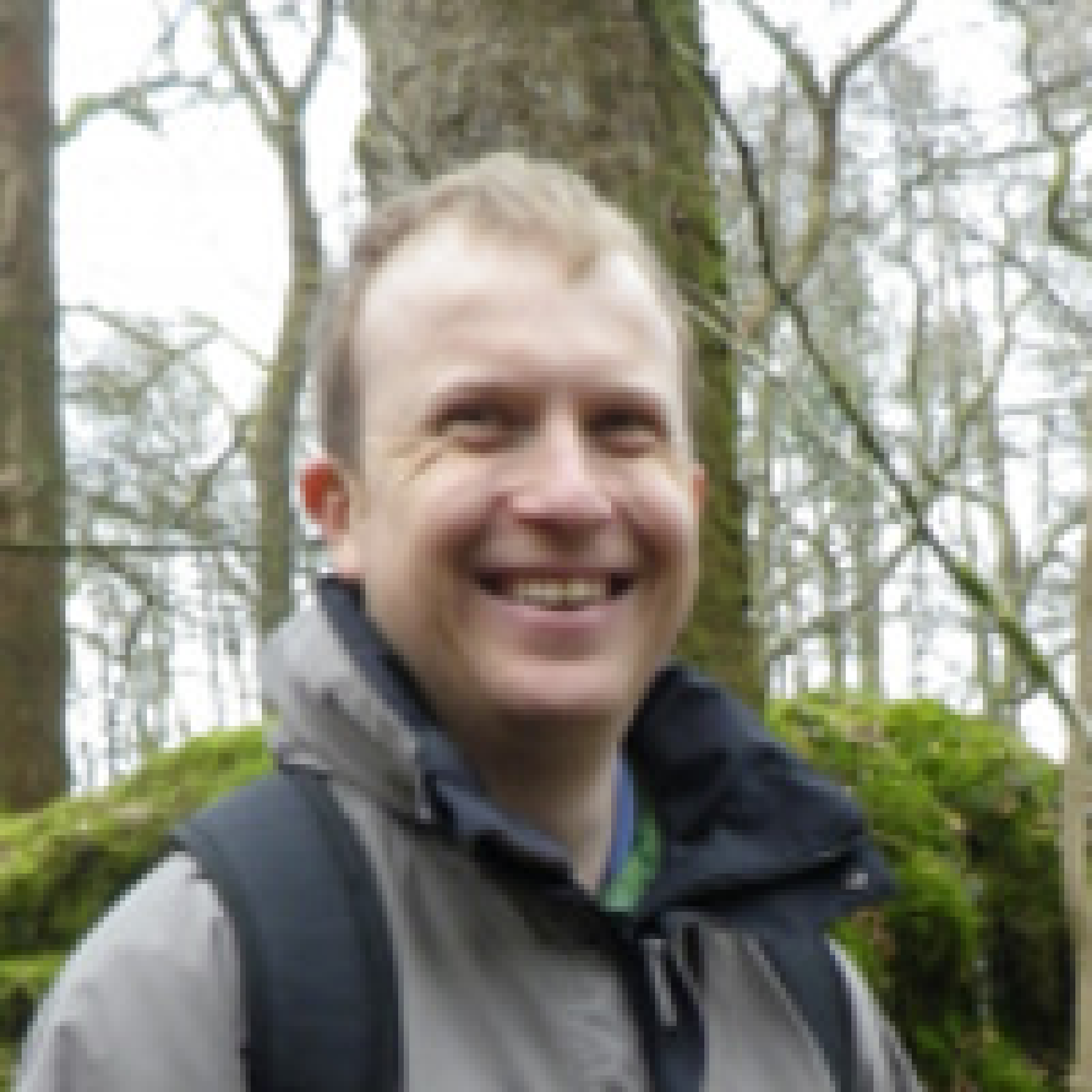}}]
{Andrew Markham} received the Bachelor's (2004) and PhD (2008) degrees in Electrical Engineering from the University of Cape Town, South Africa. 
He is currently an Associate Professor in the Department of Computer Science, at the University of Oxford, working in the Sensor Networks Group. 
His research interests include low power sensing, embedded systems and magneto-inductive techniques for positioning and communication.
\end{IEEEbiography}

\vfill
\end{document}